\documentclass{jetpl}

\usepackage{textcomp}
\usepackage{color}
\twocolumn

\lat
\usepackage{epsfig}% Include figure files
\usepackage{graphicx}% Include figure files

\title {Doping induced spin state transition in Li$_x$CoO$_2$ as studied by the GGA+DMFT calculations.}
\rtitle {Spin state transition in Li$_x$CoO$_2$}
\sodtitle {Spin state transition in Li$_x$CoO$_2$}

\author {A.\,O.\, Shorikov$^{1,2}$, V.\,V.\,Gapontsev$^{1}$, S.\,V.\,Streltsov$^{1,2}$, V.\,I.\,Anisimov$^{1,2}$}
\rauthor {A.\,O.\, Shorikov, V.\,V.\,Gapontsev, S.\,V.\,Streltsov, V.\,I.\,Anisimov}
\sodauthor {Shorikov, Gapontsev, Streltsov, Anisimov}

\address {$^{1}$M.N. Miheev Institute of Metal Physics of Ural Branch of Russian Academy of Sciences, 18, S. Kovalevskaya Street, 620990 Ekaterinburg, Russia \\~\\
$^{2}$Ural Federal University, Mira St. 19, 620002 
Ekaterinburg, Russia}

\dates {\today}{*}

\abstract{ 
Magnetic properties of Li$_x$CoO$_2$ for $x = 0.94, 0.75, 0.66$ and $0.51$ were investigated in frames of method combining Generalized Gradient Approximation with Dynamical Mean--Field Theory (GGA+DMFT). We found that a delicate interplay between Hund's exchange energy and $t_{2g}-e_g$ crystal field splitting is responsible for the high spin to low spin state transition for Co$^{4+}$ ions. The GGA+DMFT calculations show that at small doping level the Co$^{4+}$ ions adopt high spin state, while delithiation results in increase of the crystal field splitting and low spin state becomes preferable.
The Co$^{3+}$ ions were found to stays in the low spin configuration for any $x$.}

\begin{document}

\maketitle

Lithium cobalt oxide  LiCoO$_2$ is an famous material in batteries production\cite{Goodenough}.  LiCoO$_2$ exhibits no long range magnetic order down to 5K and Curie-Weiss behavior in high temperature region~\cite{Hertz08}. The local magnetic moments emerging in Li$_x$CoO$_2$ with decrease of lithium concentration $x$ (hole doping) reduces Li mobility. This can be related to the coupling between Li and magnetic Co, which leads to lowering of battery efficiency. Therefore the study of the origin of magnetism in hole doped  Li$_x$CoO$_2$ is quite important and can help in improvement of existing batteries characteristics and may suggest new ideas
in searching of novel battery materials. 

\begin {figure}
\vspace{0mm}
\includegraphics [width=0.45\textwidth]{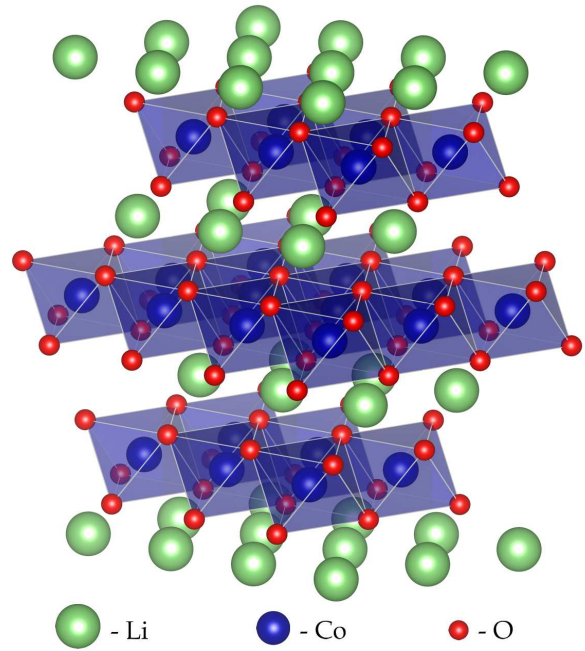}
\caption {{\bf Fig. 1. }(Color online) The crystal structure of LiCoO$_2$. Co ions shown in blue form triangular lattice. Crystal structure was plotted using VESTA software \cite{MommaK.Izumi2011}.
\label {cryst.struct}}
\end {figure}

The parent compound, LiCoO$_2$, is a quasi-two-dimensional system with the Co ions forming a triangular lattice (in $ab-$plane), see Fig.~\ref{cryst.struct}.  The Co-Co in-plane distance is two times smaller than the interplane one. The Co ions are in the CoO$_6$ octahedra, which share their edges. The Li ions are in between of the CoO$_2$ planes and donate additional $x$ electrons to these CoO$_2$ layers
in Li$_x$CoO$_2$ with $x<1$. As a result in doped material Co valence reduces from ${4+}$ ($d^5$ configuration) to ${3+}$ ($d^6$ configuration) upon changing $x$ from 0 in hypothetical CoO$_2$
to 1 in stoichiometric LiCoO$_2$. Both configurations may exist in different spin states. High spin (HS, $S=2$ for $d^6$ and $S=5/2$ for $d^5$), intermediate spin (IS, $S=1$ for $d^6$ and $S=3/2$ for $d^5$) and low spin (LS, $S=0$ for $d^6$ and $S=1/2$ for $d^5$) states.

Due to octahedral surrounding Co $3d$ band splits on $t_{2g}$ and $e_g$ sub-bands, while the trigonal distortions (due to layered structure) lead to further splitting of the $t_{2g}$ band on the higher lying $a_{1g}$ singlet and two $e_g^{\pi}$ states having lower energy. In the ionic model competition between $t_{2g}-e_g$ crystal-field splitting, $\Delta_{CFS}$, and intra-atomic Hund's rule exchange, $J_H$, defines, which spin state is stabilized in a system under consideration at given conditions\cite{Khomskii}. The transitions between different spin states (spin state transition) typically occur due to variation of temperature (as in LaCoO$_3$\cite{Imada1998}) or pressure (external, like in FeO\cite{feo}, or internal, like in RCoO$_3$\cite{Nekrasov2003}), while more exotic mechanisms such as isotope effect is possible\cite{Babushkina2014}. Moreover, doping may also trigger the spin state transition. Indeed, in Li$_x$CoO$_2$ the Co-O distance~\cite{Hertz08} shrinks with decrease of the Li content, i.e. with decrease of $x$: Co-O bond length is 1.922 \AA~for $x=1$, 1.921 \AA~ for $x=0.94$, and 1.906 \AA~ for $x=0.75$, which leads to increase of the $t_{2g}-e_{g}$ crystal field splitting and may result in the spin state transition.

Indeed, Hertz {\it et al} analyzing experimental magnetic data of Li$_{x}$CoO$_2$ proposed a spin state transition for Co$^{4+}$ ions, while Co$^{3+}$ was assumed to retain nonmagnetic LS state at any $x$\cite{Hertz08}. The magnetic susceptibility shows Curie-Weiss behavior for any doping level, which is in accord with this scenario, but effective magnetic moment, $p_{eff}$, was found to be strongly nonlinear with $x$. Hertz {\it et al} suggested change of the spin localization and onset of the two-phase region for $0.8 \lesssim x \lesssim 0.95$ to explain this feature. In addition to this scenario there are other models based of analysis of different experimental data supposing that all Co ions are in the LS state across all doping values\cite{Galakhov2006} or involving IS state\cite{Levasseur2003}. Therefore, a thorough theoretical study is needed to described evolution of Li$_x$CoO$_2$ magnetic properties with doping.

In the present paper we performed GGA+DMFT calculations of Li$_x$CoO$_2$ for various $x$. It was found that for large $x \sim 1$  Co$^{4+}$ is in the HS state, while doping results in the spin state transition to the less magnetic state. In this region Co$^{4+}$ is, indeed, mostly in the LS state, but there is a substantial contribution from the HS and IS state. The spin-state transition was shown to be induced by increase of the crystal-field splitting for $x<0.75$. 

The DFT+DMFT (called also LDA+DMFT or GGA+DMFT depending on the type
of the exchange correlation potential: LDA - local density approximation, GGA - generalized gradient approximation) approach exploits advantages of two other widely used nowadays methods: noninteracting band structure, $\varepsilon(\vec k)$, obtained within the density function theory (DFT) takes into account all peculiarities of $\varepsilon(\vec k)$ for a given material, while the dynamical mean-field theory (DMFT) takes care of many-body effects such as Coulomb correlations\cite{GGA+DMFT}. This method
was successfully used for investigation of different magnetic
phenomena including spin state transitions\cite{feo,mno,fe2o3,Streltsov14,Shorikov15}. In contrast to LDA+U or GGA+U approaches it allows not only to consider frequency dependence of the self-energy, but also simulate a paramagnetic state.  

The noninteracting GGA calculations were performed using pseudo-potential method as implemented in Quantum ESPRESSO~\cite{PW}. We used wannier function projection procedure\cite{Korotin} to extract noninteracting GGA hamiltonian $H_{GGA}$, which included both
Co $3d$ and O $2p$ states. Full many-body Hamiltonian to be solved by the GGA+DMFT is written in the form:
\begin{equation}
\hat H= \hat H_{GGA}- \hat H_{dc}+\frac{1}{2}\sum_{i,\alpha,\beta,\sigma,\sigma^{\prime}}
U^{\sigma\sigma^{\prime}}_{\alpha\beta}\hat n^{d}_{i\alpha\sigma}\hat n^{d}_{i\beta\sigma^{\prime}}.
\label{eq:ham}
\end{equation}
Here $U^{\sigma\sigma^{\prime}}_{\alpha\beta}$ is the Coulomb interaction matrix,  $\hat n^d_{i\alpha\sigma}$ is the occupation number operator for the $d$ electrons with orbitals $\alpha$ or $\beta$ and spin indexes $\sigma$ or $\sigma^{\prime}$ on the $i$-th site.  The term $\hat H_{dc}$ stands for the {\it d}-{\it d} interaction  already accounted for in the DFT, so called double-counting correction, which was chosen to be $\hat H_{dc}=\bar{U}(n_{\rm dmft}-\frac{1}{2})\hat{I}-\frac{1}{2}U_{pp}$. Here $n_{\rm dmft}$ is the self-consistent total number of {\it d} electrons obtained within the GGA+DMFT, $\bar{U}$ is the average Coulomb parameter for the {\it d} shell. We also used an additional term associated with Coulomb correlations on oxygen $2p$ shell, $U_{pp}$, which reproduces correct position of O $2p$ with respect to Co $3d$ band~\cite{Korotin00}. This additional term is responsible for the shift of fully occupied O $2p$ band downwards. In $p-d$ model it can be added directly to $p$ states in the hamiltonian, or more simply to $d$ states as a part of the double-counting term.    

The elements of $U_{\alpha\beta}^{\sigma\sigma'}$ matrix are parameterized by $U$ and $J_H$ according to procedure described in Ref.~\cite{LichtAnisZaanen}. The effective impurity problem for the DMFT was solved by the hybridization  expansion Continuous-Time Quantum Monte-Carlo method (CT-QMC) \cite{CTQMC}. Calculations for all structures were performed in the paramagnetic state at the inverse temperature $\beta=1/T$=20 eV$^{-1}$ corresponding to 580~K. Spectral functions on real energies were calculated by Maximum Entropy Method (MEM)\cite{mem}. The values of Coulomb repulsion parameter $U$ and Hund's exchange parameter $J_H$ were set to be $U$ = 7.0 eV\cite{u} and $J_H$=0.65 eV.  We used so called $U$ on oxygen correction, with $U_{pp}$ = 4 eV.   The crystal structure data for $x=0.94, 0.75, 0.66$, and $0.51$ were taken from Ref.~\cite{Hertz08}. 

\begin {figure}
\vspace{0mm}
\includegraphics [width=0.45\textwidth]{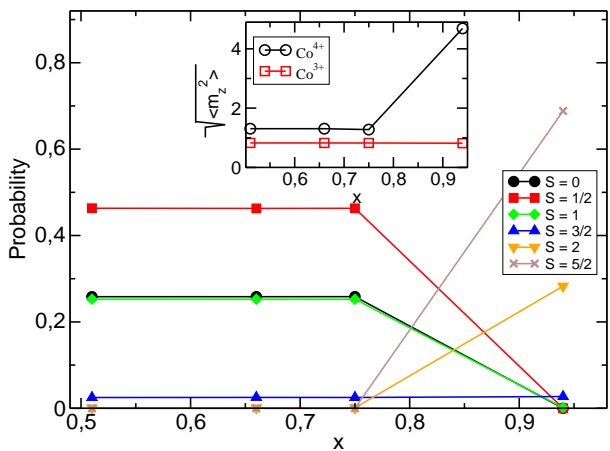}
\caption {{\bf Fig. 2. }(Color online) Probabilities of electronic configurations with different spins in the case, when all Co were supposed to have $4+$ charge state (corresponding number of electrons is $N_d+2N_p = 5 + 12 = 17$ per unit cell), as calculated in GGA+DMFT for $x=0.94,$ 0.75, 0.66, and 0.51. Inset shows local magnetic moments $\sqrt { \langle m_z^2 \rangle}$, if all Co ions are in 4+ or 3+ charge states.}
\label {fig:weight1}
\end {figure}

We performed with the calculations for two limiting cases: when (1) Co is 3+ and has $3d^6$ electronic configuration and (2) when Co is 4+ and has 5 electrons on $3d$ shell. One may see from inset of Fig.~\ref{fig:weight1} that in the first case Co$^{3+}$ turns out to be only slightly magnetic in all crystal structures. The local magnetic moments are almost the same for all structures  (0.82 $\mu_B$ for $x=0.94, 0.75$, and 0.83 $\mu_B$ for $x=0.66, 0.51$) and agree qualitatively  with conjecture of Hertz {\it et al} \cite{Hertz08}, that Co$^{3+}$ practically does not change its spin state at any doping levels. However, as opposed to the pure LS state suggested in Ref.~\cite{Hertz08} nonzero magnetic moment obtained in our GGA+DMFT calculations indicates that there is a substantial weight of the IS and HS states even for Co$^{3+}$.

In contrast to the first case of Co$^{3+}$, the second set of the calculations, corresponding to all Co having $d^5$ electronic configurations (i.e. Co$^{4+}$), demonstrates the  spin state transition for $x>0.75$. Local magnetic moment $\sqrt{\langle m_z^2 \rangle}$ for $x= 0.94$ case is 4.7 $\mu_B$, close to what one might expect for the HS state for $d^5$ configuration in an ionic model.  With doping magnetic moment drops down to 1.3 $\mu_B$.

\begin {figure}
\vspace{2mm}
\includegraphics [width=0.45\textwidth]{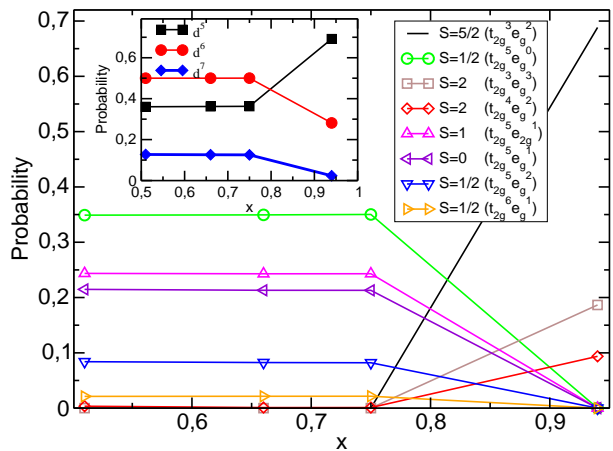}
\caption {{\bf Fig. 3. }
(Color online) The same as in Fig.~\ref{fig:weight1}, but different 
contributions to the states with the same spin are shown, e.g. there are three states with $S=1/2$ with nonzero weights ($t_{2g,\uparrow}^3,t_{2g,\downarrow}^2$, $t_{2g,\uparrow}^3 e_{g\uparrow}^1 t_{2g,\downarrow}^2 e_{g,\downarrow}^1$, and $t_{2g,\uparrow}^3 e_{g,\uparrow}^1 t_{2g,\downarrow}^3$). The inset shows relative weights of $d^5, d^6$ and $d^7$  configurations. 
}
\label {fig:weight2}
\end {figure}

This is an advantage of the CT-QMC impurity solver that it provides information about weights of different ionic configurations at given temperature.  One may see that with decrease of the doping level, $x$, and modification of the crystal structure  the  essential changes in contribution of different  electronic configurations (see Fig.~\ref{fig:weight1}) occurs. The contribution of configurations, which have the largest possible magnetic moment becomes smaller. At the same time the contribution of the IS and LS states increase. 
This is in strong contrast to the case of Co$^{3+}$ ($d^6$), where the LS state was found to be dominating for all Li concentrations (see inset in Fig.~\ref{fig:weight1}). More detailed picture of configurations probabilities is shown in Fig.~\ref{fig:weight2}.

 It is worth mentioning that in the GGA+DMFT calculation there are not only multiplets of $d^5$ electronic configuration, but also other states corresponding to, e.g., $d^6 \underline{L}$ (where $\underline{L}$ is a ligand hole) configuration. These states may have integer and zero spin; their summarized probabilities obtained are shown in the inset in Fig.~\ref{fig:weight2}. One can see that there are three most probable configurations in the case of Co$^{4+}$ for the structure corresponding to $x$=0.94: $d^5$ ($t_{2g,\uparrow}^3, e_{g,\uparrow}^2$) and $d^6 \underline{L}$ ($t_{2g,\uparrow}^3 e_{g,\uparrow}^2 e_{g\downarrow}^1$ and $t_{2g,\uparrow}^3 e_{g,\uparrow}^2 t_{2g,\downarrow}^1$), which can be treated as the HS states. The weights of these configurations drop to almost zero for $x<0.75$. For $x=0.75, 0.66$ and $0.51$ we obtained almost equal probabilities for configurations, which we treat as LS ones, namely $d^5$ ($t_{2g,\uparrow}^3 t_{2g,\downarrow}^2$), $d^6 \underline{L}$ ($t_{2g,\uparrow}^3t_{2g,\downarrow}^2e_{g,\uparrow}^1$ and $t_{2g,\uparrow}^3 t_{2g,\downarrow}^2 e_{g,\downarrow}^1$), and less probable  $d^7 \underline{L}^2$ ($t_{2g,\uparrow}^3 e_{g\uparrow}^1 t_{2g,\downarrow}^2 e_{g,\downarrow}^1$ and $t_{2g,\uparrow}^3 e_{g,\uparrow}^1 t_{2g,\downarrow}^3$). All other configurations have probabilities of order of $10^{-3}$ or smaller.

In order to get insight into the nature of the spin state transition we 
estimated values of the $t_{2g}-e_g$ crystal field splitting for different $x$ as difference between centers of gravity of corresponding partial density of states (DOS) in the GGA\cite{Streltsov2005}: $\Delta_{CFS}=2.28$ eV for $x=0.94$, $\Delta_{CFS}=2.32$ eV for $x=0.75$, $\Delta_{CFS}=2.38$ eV for $x=0.66$ and $\Delta_{CFS}=2.40$ eV for $x=0.51$. Gradual increase of the crystal-field splitting is related to decrease of Co-O bond distance as it was discussed above and this increase in its turn leads to the spin state transition. 

One may use a pure ionic model for the qualitative analysis of the spin state transition. In the case of $d^6$ configuration $\Delta_{CFS}$ competes with $2J_H$ to suppress the HS state, while for $d^5$ configuration $3J_H$ act against the crystal field splitting (see, e.g.,~\cite{Streltsov2011}). This is the reason why the spin state transition occurs for Co$^{4+}$, but not for Co$^{3+}$.

\begin {figure}
\vspace{2mm}
\includegraphics [width=0.45\textwidth]{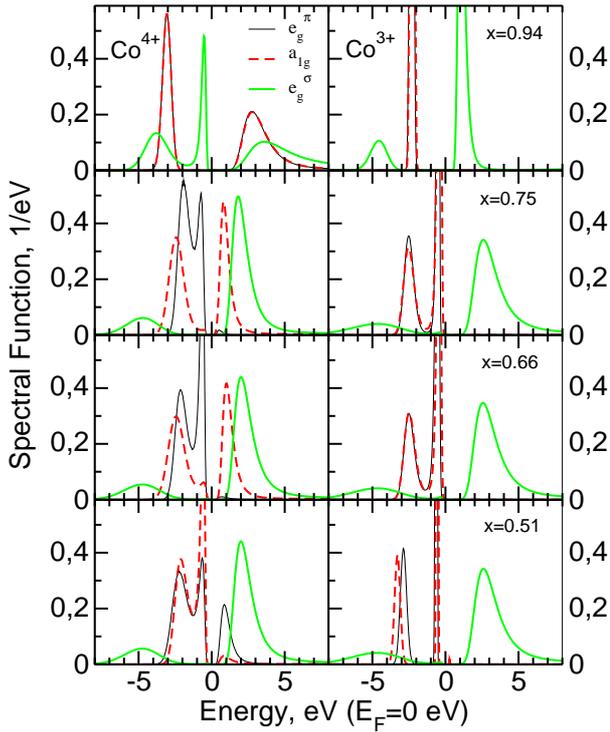}
\caption {{\bf Fig. 4. }(Color online) Evolution of the GGA+DMFT spectral function for Co$^{4+}$ (left panel) and Co$^{3+}$ (right panel) in Li$_x$CoO$_2$.}
\label {fig:spec_funct_real}
\end {figure}

In Fig.~\ref{fig:spec_funct_real} we present spectral functions, $A(\omega)$, which were obtained within GGA+DMFT calculations for different $x$. One may see that while $A(\omega)$ practically does not change with $x$ for Co$^{3+}$. This is mainly due to absence of the spin state transition for this Co. In the case of $x=0.94$ and Co$^{3+}$ we obtain a gap about 2.8 eV (here and below we measure the gap at the half maximum of corresponding peaks), which agrees with 
experimental value 2.7~\cite{vanElp91} for pure LiCoO$_2$. 

In contrast to Co$^{3+}$ there is a drastic change of the band gap in case of Co$^{4+}$, which is  $\sim 2.5$ eV  for $x=0.94$, but almost disappears for smaller $x$. This is related with a very different effective Coulomb repulsion $U_{eff}$ for the HS and LS states. Indeed, if one would recalculate $U_{eff}$ in ionic approximation using a standard definition
\begin{eqnarray}
U_{eff} = E(d^{n+1}) + E(d^{n-1}) - 2 E(d^n),
\end{eqnarray}
where $E$ is the total energy of corresponding configuration, when it turns out that for the HS state $U_{eff} = U + 4 J_H$ and
for the LS state $U_{eff} = U - J_H$. Such a large difference will be greatly reduced by band effects, but is still noticeable in our GGA+DMFT results.

In conclusion our calculations carried out within the DFT+DMFT method show that there is the high spin to low spin transition Li$_x$CoO$_2$ with decrease of lithium concentration. The nature of this transition is a delicate balance between crystal splitting and Hund's rule interaction, which is tilted with decrease of the doping level due to change of the crystal structure. The decrease of the Co-O distance results in increase of crystal field splitting that makes the LS configuration of both Co$^{3+}$ and Co$^{4+}$ more  preferable. Thus, reduction of the Co-O bond length can stabilize the LS state in Li$_x$CoO$_2$ and hence can help to avoid appearance of magnetic traps, which improves ionic conductivity. This idea can be used in fabrication of Li$_x$CoO$_2$ thin films batteries by appropriate choice of a substrate.   

This work was supported by the grant of the Russian Scientific Foundation (project no. 14-22-00004).


\begin{thebibliography}{99}
\bibitem{Goodenough} K. Mizushima {\it et al}, Mat. Res. Bull. \textbf {15}, 783 (1980).
\bibitem{Hertz08} J. T. Hertz {\it et al}, Phys. Rev. B \textbf {77}, 075119 (2008).
\bibitem {MommaK.Izumi2011} K. Momma and F. Izumi, 
J. Appl. Crystallogr., \textbf {44}, 1272-1276 (2011).
\bibitem{Khomskii}  D. I. Khomskii, Transition Metal Compounds (Cambridge University Press, 2014).
\bibitem{Imada1998} M. Imada, A. Fujimori, and Y. Tokura, Rev. Mod. Phys. \textbf {70}, 1039 (1998).
\bibitem{feo}	A.O. Shorikov {\it et al}, 
%Z.V. Pchelkina,  V. I. Anisimov, S.L. Skornyakov, M. A. Korotin
Phys. Rev. B \textbf {82}, 195101 (2010).
\bibitem{Nekrasov2003} I. A. Nekrasov {\it et al}, Phys. Rev. B \textbf {68}, 235113 (2003).
\bibitem{Babushkina2014} N. A. Babushkina {\it et al}, JETP \textbf {118}, 266 (2014).
\bibitem{Galakhov2006} D. G. Kellerman {\it et al}, Phys. Solid State \textbf {48}, 548 (2006).
\bibitem{Levasseur2003} S. Levasseur {\it et al}, Chem. Mater. \textbf {15}, 348 (2003).
\bibitem {GGA+DMFT} V. I. Anisimov  {\it et al.}, J. Phys.: Condens. Matter \textbf {9}, 7359
(1997); A. I. Lichtenstein and M. I. Katsnelson, Phys. Rev. B \textbf {57}, 6884 (1998); K. Held {\it et al.}, Phys. Stat. Sol. (b) \textbf {243}, 2599 (2006).
\bibitem{mno}	Jan Kunes {\it et al}, Nature Materials \textbf {7}, 198 (2008).
\bibitem{fe2o3}	J. Kunes {\it et al}, Phys. Rev. Lett. \textbf {102}, 146402 (2009).
\bibitem{Streltsov14} S. V Streltsov and D. I. Khomskii, Phys. Rev. B \textbf {89}, 161112 (2014).
\bibitem{Shorikov15} A. O. Shorikov, A. V. Lukoyanov, and V. I. Anisimov and S. Y. Savrasov, Phys. Rev. B \textbf {92}, 035125 (2015)
\bibitem{PW} P. Giannozzi {\it et al}, J. Phys. Condens. Matter 21, 395502 (2009).
\bibitem{Korotin} Dm. Korotin {\it et al}, Euro. Phys. J. B {\bf 65}, 1434 (2008).
\bibitem{Korotin00} M. Korotin, T. Fujiwara, and V. Anisimov, Phys. Rev. B {\bf 62}, 5696 (2000).
\bibitem{LichtAnisZaanen} A.~I.~Liechtenstein, V.~I.~Anisimov, J.~Zaanen, Phys. Rev. B {\bf 52}, R5467 (1995).
\bibitem {CTQMC} P. Werner {\it et al}.,  Phys. Rev. Lett. {\bf 97}, 076405 (2006).
\bibitem{mem} M.~Jarrell, J.~E.~Gubernatis, Phys. Rep. {\bf 269}, 133 (1996).
\bibitem{u} M.M. Markina {\it et al}., Phys. Rev. B {\bf 89}, 104409 (2014). 
\bibitem{Streltsov2005} S. V. Streltsov {\it et al}. Phys. Rev. B \textbf {71}, 245114 (2005).
\bibitem{Streltsov2011} S. V. Streltsov and N. A. Skorikov, Phys. Rev. B  \textbf {83}, 214407 (2011).
\bibitem{vanElp91} J. van Elp {\it et al}, Phys. Rev. B {\bf 44}, 6090 (1991).
\end{thebibliography}
\end {document}